\documentclass{sig-alternate-10pt}

\usepackage[ruled]{algorithm2e}

\SetAlFnt{\small}
\SetAlCapFnt{\small}
\SetAlCapNameFnt{\small}
\SetAlCapHSkip{0pt}
\IncMargin{-\parindent}

\usepackage{amsmath}
\usepackage{multirow}
\usepackage{rotating}
\usepackage{tabularx}
\usepackage{caption}
\usepackage{booktabs}
\usepackage{pgf}
\usepackage{tikz}
\usepackage{pifont}
\usepackage{xcolor}
\usepackage{ctable}
\usepackage{url}
\usepackage{wasysym}

\usepackage{hyperref}
\hypersetup{
    pdftitle={Sybil Attacks in Peer to Peer Overlays},
    pdfsubject={survey},
    pdfauthor={The authors},
    pdfkeywords={survey},
    pdfpagemode=pagewidth,
    plainpages=false,
    colorlinks,
    urlcolor=blue,	
    linkcolor=red,    
    citecolor=purple,  
    bookmarksnumbered
}

\colorlet{darkblue}{blue!50!black}
\colorlet{darkred}{red!70!black}
\colorlet{darkgreen}{green!50!black}

\usetikzlibrary{shapes,snakes,calendar,matrix,backgrounds,folding}

\newcommand{\handout}[5]{
   \renewcommand{\thepage}{#1-\arabic{page}}
   \noindent
   \begin{center}
   \framebox{
      \vbox{
    \hbox to 5.78in { {\bf CS 388H Introduction to Cryptography} \hfill #2 }
       \vspace{4mm}
       \hbox to 5.78in { {\Large \hfill #5  \hfill} }
       \vspace{2mm}
       \hbox to 5.78in { {\it #3 \hfill #4} }
      }
   }
   \end{center}
   \vspace*{4mm}
}

\begin{document}
\title{The Sybil Attacks and Defenses: A Survey\titlenote{Also appears as an invited article in Smart Computing Review, vol. 3, no. 6 pp 480-489, 2013.}}
\numberofauthors{2}
\author{
\alignauthor
Aziz Mohaisen\\
\affaddr{VeriSign Labs}\\
\affaddr{Reston, VA 20190, USA}\\
\affaddr{{amohaisen@verisign.com}}\\
\and
Joongheon Kim\\
\affaddr{University of Southern California}\\
\affaddr{Los Angeles, CA, USA}\\
\affaddr{{joongheon.kim@usc.edu}}\\
}
\conferenceinfo{ICUIMC'11,} {February 21--23, 2011, Seoul, Korea.}
\CopyrightYear{2011}
\crdata{978-1-4503-0571-6}
\clubpenalty=10000
\widowpenalty = 10000
\maketitle

\begin{abstract}

In this paper we have a close look at the Sybil attack and advances in defending against it, with particular emphasis on the recent work. We identify three major veins of literature work to defend against the attack: using trusted certification, using resources testing, and using social networks. The first vein of literature considers defending against the attack using trusted certification, which is done by either centralized certification or distributed certification using cryptographic primitives that can replace the centralized certification entity. The second vein of literature considers defending against the attack by resources testing, which can by in the form of IP testing, network coordinates, recurring cost as by requiring clients to solve puzzles. The third and last vein of literature is by mitigating the attack combining social networks used as bootstrapping security and tools from random walk theory that have shown to be effective in defending against the attack under certain assumptions. Our survey and analyses of the different schemes in the three veins of literature show several shortcomings which form several interesting directions and research questions worthy of investigation.
\end{abstract}


\section{Introduction}\label{sec:intro}
The peer-to-peer paradigm of computing has a lot of advtnages over other conventional paradigms. For example, in this paradigm resources such as bandwidth, memory, and data are made available to other all participating users~\cite{Danezis05sybil}. Broadly, this paradigm includes structured and unstructured systems. Structured overlays, such as Kademlia \cite{MaymounkovM02} and Chord \cite{StoicaMKKB01}, provide deterministic mechanisms for data and peers discovery while unstructured overlays, such as Gnutella \cite{WangYL07}, organize peers in a random graph and use flooding for peers and data discovery. Most of the popular peer-to-peer systems lack ``centralized authorities'' which makes this paradigm robust against failure attacks. On the other hand, the lack of such centralized authority leads to so many challenging security issues: most security services necessary for securing networked systems require a type or another of a centralized authority making these services unavailable for the peer-to-peer systems~\cite{PathakI06}. Even worse, the fully decentralized and open nature of many of these systems enable a wide range of security threats unknown in other distributed systems, including the {\em Sybil attack} \cite{Douceur02thesybil}.

The Sybil attack is well-known in the context of peer-to-peer, wired, and wireless networks. In its basic form, a peer representing the attacker generates as many identities as she can and acts as if she is multiple peers in the system~\cite{Douceur02thesybil} aiming at disturbing the normal behavior of the system. The number of identities that an attacker can generate depends solely on the attacker's capabilities which are limited by the bandwidth required for responding to concurrent requests by other peers in the system, the memory required for storing routing information of other peers corresponding to each and every generated Sybil identity, and computation resources required for serving concurrent requests without noticeable delay. With the sharp hardware growth (e.g., in terms of storage capacity and processing) as well as the wide spread of broadband Internet with high bandwidth rates, even attackers running on `commodity'' hardware can cause a substantial harm to large systems.

The attack itself is popular and effective in many contexts and on my services that are essential in peer-to-peer systems as well as other generic distributed systems and paradigms. Such contexts include  voting systems reputation systems, routing, distributed storage, among others. To illustrate how this attack works in real systems, imagine a recommender system built over a peer-to-peer overlay~\cite{YuSKGX09}. In such system, the goal is to filter information that are likely to be of interest to users based on others' recommendations. In that context an attacker that can act as multiple users by faking multiple identities (Sybil) can easily out-vote legitimate users' votes on legitimate objects subject of voting. This is almost guaranteed given that the number of legitimate users that normally vote on objects are always not more than $1\%$ of overall users in the system in any realistic recommendation system~\cite{YuSKGX09}.  Such an attack becomes appealing to attackers, who are potentially users in the system trying to take advantage of the system operation given high incentives. For example, many online market-places, such as eBay, use recommendations of customers to determine the reputation of their sellers, people who use their platform to sell goods, and thus there is an appealing incentive for such sellers to misbehave to gain higher reputation. The same scenario arises in many other contexts such as peer-to-peer file sharing where contents are rated by users, bandwidth is assigned based on reputation, or even that reputation is used to determine the goodness of contents distributed by users. In all of such examples, incentives exist for users to misbehave and the Sybil attack is proven to be a powerful tool for such attacker to achieve his goals.

To defend against the attack, there has been several attemtps in the form of defenses, or mitigations, to defend against or limit the impact of the attack. Such attentps can be classified broadly into two schools of thoughts: centralized and decentralized (i.e., distributed) defenses. In centralized defenses~\cite{Douceur02thesybil,CastroDGRW02,AdyaBCCCDHLTW02,LedlieS05}, a centralized authority is made responsible for verifying the identity of each and every users in the systems. While this defense is somewhat effective in defending against the attack, it makes certain assumptions about the system some of which are not easy to achieve in peer-to-peer decentralized systems. First of all, and as the name and the description of operation tells, such systems require a centralized authority, that in many of such systems may not be affordable for both security and functionality reasons. Also, even if such centralized authority existed, it requires some credentials related to the users in the system so as to match these credentials of the users to their digital identity: in many settings, obtaining such credentials is very challenging.

On the other hand, decentralized defenses including, but not limited to, the work in ~\cite{PKIP2P,LesueurMT08a,LesueurMT08b,LesueurMT08c,AvramidisKD07,borisovPuzzle,YuGK07,YuGKX08,YuKGF06,YuKGF08,YuSKGX09,gatekeeper-infocom11,lassK10,lesniewski2008sybil} do not require such authorities in the system and are well-designed for decentralized peer-to-peer systems. At the core of their operation, such defenses weigh collaboration among users in the system to admit or reject users, who are potentially attackers. The admision or rejection of users is based on credentials associated with users, as it is the case of cryptographic distributed defenses, or network properties of legitimate honest users, as it is the case of Sybil defenses using social graph. In either of the solutions, the ultimate goal of the defenses is to simulate the power of the centralized authority in a decentralized manner and use such power to detect Sybil and honest nodes.

Another classification of defenses could be according to the way such defenses operate. Accordingly, existing defenses in literature can be classified into defenses using trusted certification---in which certificates are typically generated for honest users and verified against a public key of a trusted authority, incurring cost---in which users penalized by some cost so that limit them to their amount of available resources and thus reduce their misbehavior, and social networks based Sybil defenses.

Such defenses greatly differ in their assumptions, the type of networks they are applied to, the guarantees they provide, and the cost they incur. To this end, this, this paper this paper is dedicated for reviewing, summarizing, comparing, and showing shortcomings of the existing literature on such defenses. Our method in this survey is characterized by two aspects: first, we review each category of each work direction the defenses and show their merits in defending against the attack in the claimed context. Second, we summarize the direction by showing the main shortcomings that lead to open problems worthy of investigation. This latter part shed the light on that the technical contribution toward solving the problem is fragile and that a lot of investigation is required in order to solve the problem.

To this end, we summarize the organization of the rest of this paper. In section~\ref{sec:prelim} we introduce preliminaries, including a functional classification of defenses. In section~\ref{sec:trusted_certification} we review the trusted certification based category of defenses while in section~\ref{sec:resources_testing} we introduces the category of resources testing followed by social network-based defenses in section~\ref{sec:graph}. In section~\ref{sec:related} followed by concluding remarks in section~\ref{sec:conclusion}.

\section{Model, settings, and objectives}\label{sec:prelim}
In this section we elaborate on the problem in hand and state the attacker model conventionally assumed in most of Sybil attack studies and defenses. We further explore the objectives of the attacker and the objective of the defenses proposed in literature.

\subsection{Problem Statement and Model}\label{sec:problem_statement}

The problem is stated as the ability of a single user in the system to act as if she is multiple users with different identities. This is problematic for so many applications since the correctness of such applications depends on the behavior of peers, their numbers, and willingness to participate collaborate honestly in the system. However, such goal cannot be satisfied  with Sybil identities of a single attacker which tries to bias the overall behavior of the system.

The attacker is formally characterized by the number of fake identities that she can inject in the system. The attackers incentive is to maximize this number. The value and meaning of the number of identities generated by the attacker and injected into the system depends on the application itself and varies from an application to another. For example, to attack a recommendation system, it is enough to have a matching sum of $1\%$ of the honest users in the system as fake identities. This is particularly enough to bias the behavior of the system and outvote the honest nodes in the system since it is generally observed that, even for the very popular objects in this system, only $1\%$ of the honest node vote for it. This is, by having a single identity more than the exact number of honest nodes voting on an object in the system would enable the Sybil attacker to outvote the honest nodes.

On the other hand, in other systems such as the mixing networks used for communication anonymization (e.g., Tor network) sufficiently small number of Sybil identities may present a serious breach to the guarantees in the System. Theoretically, the compromise of two nodes on a circuit  is sufficient to identify the sender and the receiver of the communication on such mix network~\cite{SyversonGR97}. On the other hand, the compromise of sufficient large number of identities in the network would enable the attacker from monitoring arbitrary number of circuits. Other applications where the number of identities itself matters include attacks on file sharing systems~\cite{WangKAD}, among many others.

To sum up, the attacker's objective is to maximize the number of Sybil identities in the overlay though in a few a small number of Sybil identities  suffices to thwart the application in some cases.

\begin{figure}[htbp]
\begin{center}
\begin{minipage}[t]{0.49\textwidth}
\begin{center}
\begin{tikzpicture}
[level distance=10mm,
every node/.style={inner sep=1pt},
level 1/.style={sibling distance=40mm,nodes={fill=blue!10}},
level 2/.style={sibling distance=15mm,nodes={fill=blue!10}},
edge from parent/.style={draw,black,thick}]
\tikzstyle{every node}=[draw,shape= rectangle];
\node {\bf Sybil Defenses}
		child {node {Certification}
			child{node {C/CA}}
			child{node {D/C}}
			child{node {T/D}}
		}
		child {node {Resources Testing}
			child{node {IP/T}}
			child{node {R/C}}
			child{node {S/G}}
		};
\end{tikzpicture}
\end{center}
\end{minipage}
\end{center}
\caption{An illustration of the different types defense for Sybil attack in P2P overlays. C/CA stands for centralized certificate authority, D/C for decentralized cryptographic, T/D for trusted devices, IP/T for IP testing, R/C for recurring cost, and S/G stands for social graph-based approaches.}\label{fig:deftree}
\end{figure}
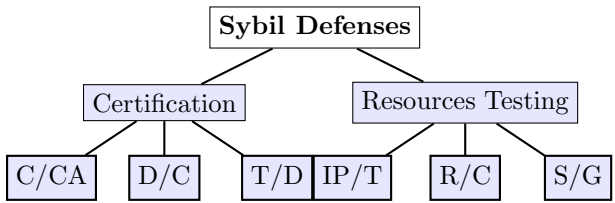

\subsection{Defense Goals and success metric}\label{sec:solution_objectives}
The ultimate objective of Sybil defenses is to eliminate the Sybil attack by detecting and isolating Sybil identities, or peers that generate such identities, from the overlay. However, this ultimate goals is not always possible due to that most defenses, except of the centralized trusted certification based scheme, have false positive and false negative in their detection mechanism that might tolerate some Sybil nodes while mistakenly report other honest nodes as observed in the false positive and false negative errors. In false negative, Sybil nodes are reported to be honest nodes. On the other hand, in false negative honest nodes are reported to be Sybil nodes. The realistic and practically attainable goal of defense mechanisms is to minimize false negative rate as much as possible.

\subsection{Deep of Sybil defenses}

Beside the broad classification of Sybil defenses into centralized and decentralized defenses as shown in section~\ref{sec:intro}, the defenses we survey in this paper include two major and broad categories of techniques: trusted certification and resource testing categories as shown in Figure~\ref{fig:deftree}. Among the schemes in the trusted certification category, we survey works that use centralized certification authority, decentralized cryptographic primitives, or trusted devices. Among works that use resources testing, we are particularly interested in works that use IP testing, cost recurring, and social graphs. While broad survey is provided for these techniques, detailed descriptions are provided for cryptographic primitives and social graph-based techniques.

\section{Using Trusted Certification}\label{sec:trusted_certification}
The trusted certification approach is arguably the most popular method in the context of this study since it has been proven by Douceur \cite{Douceur02thesybil} for its potential to {\em eliminate} the Sybil attack \cite{levine2006survey}. In the conventional form of this approach, a centralized authority (CA) is used to ensure that the identities assigned to each peer are unique and legitimate by matching these identities to pre-assigned credentials. These credentials may include cryptographic keys, synchronized random strings which are usually generated by  one time password generators -- OTP, or digital certificates issued by the centralized authority.

While the aforementioned conventional form of centralized certification authority is well-defined in literature, there has been some efforts to define a distributed certification schemes by applying cryptographic primitives suitable for distributed multiparty models which enable collaboration among {\em supposedly} honest peers to certificate other peers joining the overlay. In this section we review both approaches. In particular, we report some of the works in literature on centralized certification authorities and provide details for cryptographic primitives to enable distributed certification systems.

\subsection{Centralized Certification Authority}
Centralized trusted certification is potentially the only method that can eliminate the Sybil attack \cite{Douceur02thesybil}. There has been some works on the usage of centralized certification authorities for credentials generation, assignment, and verification in the context of P2P overlays. For instance, works that use social graphs which are explained in section \ref{sec:graph} and utilize public key cryptography as a building block assume the authenticity of public keys of users via certificates assigned to users via a centralized authority in an offline phase \cite{YuKGF06}. Other examples of schemes that utilize CCA-based approach include the works in \cite{CastroDGRW02}, \cite{AdyaBCCCDHLTW02}, \cite{Douceur02thesybil}, and \cite{LedlieS05}.

\subsection{Cryptographic primitives}\label{sec:crypto}

Recently, a few works \cite{PKIP2P,LesueurMT08a,LesueurMT08b,LesueurMT08c,AvramidisKD07} on cryptographic primitives have been proposed. These primitives aim at providing an infrastructure for authenticating peers in order to make the Sybil attack harder to apply by having only legitimate peers participating in the overlay. Generally, these works try to exploit public key infrastructure (PKI) in a distributed manner and using {\em threshold} cryptographic ingredients (e.g., secret sharing and threshold signatures) in order to ensure a collaboration among {\em supposedly} honest users to authenticate peers that join the overlay over its operation time. Interestingly, the explicitly stated motivation beyond some of these primitives (e.g., \cite{PKIP2P,LesueurMT08a,LesueurMT08b,LesueurMT08c}) is that many of the non-cryptographic protocols in lietrature assume the existence of a certification system for legitimate users in the overlay (e.g., SybilGuard and SybilLimit). Hence, the cryptographic approaches are designed to ensure successful operation of such protocols. In this section we review two recent cryptographic approaches -- variations of the work in \cite{PKIP2P}  have also appeared in \cite{LesueurMT08a,LesueurMT08b,LesueurMT08c}.

\subsection{Trusted Devices}\label{sec:trusted_devices}
	Similar to the idea of trusted certification, some works have suggested the usage of trusted devices or trusted modules that store certifies, keys, or authentication strings previously assigned to users by a centralized authorities. Such devices are basically hard to obtain for their potentially high price and hence can be used for limiting the chances for Sybil attackers. Examples of such mechanism are proposed in \cite{RodrigruesLS02} and \cite{NewsomeSSP04}, though the later work is on wireless sensor networks.  In theory, when the intent of the attacker to obtain as much as possible of Sybil identities, these defenses might be effective. However, in cases such as anonymity (Tor for instance) and recommender systems, given that a fewer Sybil identities could cause a great harm, these defenses are obsolete.

\section{Resources Testing}\label{sec:resources_testing}
	The basic idea beyond the resources testing approach for defending against Sybil attack is to check if a set of identities associated with supposedly different users own enough resources that match the number of identities. These resources may include computation power, bandwidth, memory, IP addresses, or even trust credentials. Though the idea of resources testing is shown by Douceur for its ineffectiveness \cite{Douceur02thesybil}, some researchers have argued it to be a minimal defense. This is, the method is not supposed to entirely eliminate the attack but to make it harder to apply.

	In theory, the majority of schemes in this category of defenses limit the number of Sybil identities to smaller number than in the scenario without defenses in place. However, in practice, even smaller number of Sybil identities are enough to thwart the availability and security of many systems. For instance, as mentioned before, anonymity in anonymous systems such as Tor depends on two nodes per circuit. Also, it is enough to have $1\%$ of fake identities in online reputation systems in order to out-vote legitimate nodes. In sipte of that, we review some of the works that use the resources testing approach and show their shortcomings though we should keep in mind that these schemes are mitigations (i.e., discourage the attacker) rather than eliminating the attack.

\subsection{IP Testing}
	Generic testing schemes include the testing of IP address of peers trying to determine their location and match that to their activities. In particular, if an amount of activities is generated from the same particular geographical area, it is likely that some of these activities are due to sybil identities. Beside, the assumption in such works is that it is not cheap to obtain IP addresses in different geographical areas. For example, Freedman et al. \cite{FreedmanM02} introduced Tarzan in which IP addresses of peers are tested based on their geographical location in a particular autonomous system. Similar results are introduced by Cornelli et al in \cite{CornelliDVPS02} and \cite{CornelliDVPS02}.

	The main assumption for these works is that IP addresses are hard to obtain in geographically wide areas. However, with recent indicators for the existence of gigantic Botnets \cite{KangCLTKNWSHDK09}, compromised hosts under control by a single administrative entity and reside in different autonomous systems, it is quite certain that such defense mechanisms are useless.

\subsection{Recurring Cost}\label{sec:rec_cost}
	Some works have suggested recurring costs as a method of defending against the sybil attack. In particular, computational puzzles \cite{borisovPuzzle,li2012sybilcontrol} and Turing tests (e.g., CAPTCHA \cite{AhnBHL03}) are suggested as solutions. However, for the same reason that IP testing would not work against an attacker that controls a Botnet, these cost-based scheme will not work as well. Furthermore, for CAPTCHA-like solutions, it has been shown that the Sybil attacker may post the CAPTCHA tests on sites controlled by her for users who may solve the test for getting access to the service provided by the attacker. Also, some versions of CAPTCHA are vulnerable to some image processing attack \cite{YanA07}.

\section{Social network-based defenses}\label{sec:graph}
      While most of the previously proposed solutions to the problem of Sybil attack in distributed systems have limitations and shortcomings, in a way or another, social-network based Sybil defenses try to overcome such shortcomings in several elegant ways. First, social network-based Sybil defenses are mostly decentralized solution to the problem of Sybil attack, which means that these designs operate without any centralized authority---a feature that is highly desirable in necessary in most distributed systems. This decentralized model of operation is further made easier thanks to the random walk theory, an ingredient mostly utilized in these defenses. Second, these defenses utilize the trust of social links among social nodes, making collaboration among honest nodes possible and easy. Third and last, these defenses are shown in several studies (citation?) to be practical and effective in defending against Sybil defenses at low cost, and are further developed as components in many services, including distributed hash tables (DHT), Sybil-resistant voting, and are utilized in mobile networks routing.

      Although they differ greatly in their design details and operation, all social network-based Sybil defenses have two common assumptions: an algorithmic property, called the fast mixing property, and trust. First, these defenses are based on the ``fast mixing'' property of social graphs (a property that we formally define below). Informally, the fast mixing property of the social graph implies that the ``honest'' nodes in such graph are well-enmeshed and the honest region does not contain a sparse-cut---a cut that connects two large subsets of honest nodes with a few social links. For simplicity, the fast mixing property of social graphs implies that a random walk from any arbitrary node in the social graph would reach very close to the stationary distribution of the Markov Chain (MC) defined on that graph after a few walk steps. Such number of steps is suggested to be $10$ to $15$ steps in a network of million of nodes.

      The second assumption common to this vein of defenses is trust. In particular, all of these defenses assume a good trust value in the underlying social graphs as indicated, for example, by face-to-face interactions among the nodes. This particular assumption is necessary in order to reason about the hardness of infiltrating the social network by arbitrarily many attackers' social links. While the operation of the Sybil defense for correctly identifying ``honest'' nodes in the social graph is guaranteed by the fast mixing assumption, and the construction of the corresponding scheme that uses such algorithmic property, the power of identifying Sybil nodes is only guaranteed assuming that the attacker, or attackers collectivity, control a few links between themselves and other honest nodes in the social-graph (such links are called attack edges).

      In the following we review some of the widely cited works on social network-based Sybil defenses. The reader is referred to compressive works in this regard, and to the following related works on the matter.

\subsection{SybilGuard}
    The design of SybilGuard, due to Yu et al.~\cite{YuKGF06,YuKGF08}, uses the fast mixing property of trust-possessing social networks to detect Sybil nodes. Technically, SybilGuard consists of two phases: initialization and online detection phases. In the initialization phase, each node constructs its routing table, as random permutation of its adjacent nodes for pairs of incoming and outgoing edges. Next, each node initiates a random walk of length $w=O(\sqrt{n}\log n)$ and propagates it to its adjacent nodes following the routing tables constructed using the random permutation. Each node at the path of the random walk registers the the public of the random walk originator, and later acts as a witness of that node when that node is a suspect. Furthermore, using the back-traceability of the random walks, each originator of a random walk receives the list of ``witnesses'' (i.e., the nodes which register the originator's public key and lie on the path constructed by the random walk of the originator).

    In the online phase a verifier determines whether a suspect is honest or not as follows. First, the suspect sends the addresses of the ``witnesses'' on his random route to verifier. Accordingly, verifier compares the list of witness to his list of verifier route. If no intersection among the two sets (an event that has a very probability) the verifier aborts and rejects the suspect. Otherwise, the verifier continues by requesting the nodes on the intersection among the two sets to verify if the suspect has a public key registered with them. If the suspect is verified by the intersection nodes, the verifier accepts the suspect or mark it as a Sybil node otherwise.

\subsection{SybilLimit}
    Unlike SybilGuard in which a single long random walk is used, SybilLimit suggests the use of several shorter random walks. Also, unlike SybilGuard where public keys of verifiers and suspects are registered on nodes in the social graph, SybilLimit~\cite{YuGKX08} suggests the registration of such keys on edges in the social graphs. SybilLimit consists of an initialization phase and an online verification phase. In the initialization phase, each node constructs its routing table using the same method described in SybilGuard and performs $r=O(\sqrt{n})$ random walks each of length $w=O(\log n)$ where $O(\log n)$ is the mixing time of the social graph---which is used as $10$ to $15$ in a million of nodes social graph, and is theoretically assumed to be sufficient to sample nodes from a distribution that is very close to the statistical distribution. Unlike SybilGuard where all nodes on the path of the random route are used for registering the public key of the originator of the random walk, the last {\em edge} in each walk among the $r$ random walks is used for registering the public key of that originator node (each of such edges is called {\em tail}). In particular, the public key of the originator of the walk is registered at the last node in the walk under the last edge through which the random walk has arrived. Also using the back-tracebility property of the random routes, the witnesses which register the the public key of the originator node (which could be either a suspect or a verifier) return their identities to that node. The same process is performed by every node in the social graph and sets of witnesses (or verification nodes) are collected by each node that originates the random walks of registration.

    In the online phase, also same as SybilGuard, the suspect sends the identifiers and addresses of the witnesses to the verifier node which compares the witnesses in the suspect's list trying to find a collision. If a collision happens in the two sets at the verifier side, the verifier asks the witness with common identity in both sets to verify the identity of the suspect and decides whether to accept or reject the suspect based on the outcome of this process. If no intersection happens between the two sets (which has a very small probably) the verifier aborts and rejects the suspect, by labeling her as an attacker.

    The main ingredients used for reasoning about the provable guarantees of the SybilLimit are same as those in SybilGuard. In particular, given that the random walk length $w$ is the mixing time of the social graph, the last node selected in such random walk is according to the stationary distribution. Furthermore, the last edge in the random walk is selected ``almost'' uniformly at random from the edges in the social graph. Also, given that $r=O(\sqrt{n})$, an intersection between the sampled edges of the verifier and the suspect exists with an overwhemling probability, if the hidden constanct $r_0$ (where $r=r_0\sqrt{n}$) is choosen correctly. The authors refer to this condition as the ``intersection'' condition, which is used for ensuring a high probability for intersection of random walks by nodes in the honest region.     As in SybilGuard, assuming $g$ attacker edges, the attacker is allowed to register his public keys of Sybil identities on at most $gwr = O(g\sqrt{n}\log n)$ tails (called tainted tails). In such case, each attached edge introduces additional $O(\log n)$ Sybil identities (assuming that the attacker uses the optimal attack strategy by registering different public keys of different Sybil identities at each possible tainted tail).

    SybilLimit also greatly depends on $w$ for its security. Since there is no mechanism for estimating the exact value of the parameter, undersetimating or overestimating such parameter are both problematic as shown above. SybilLimit also provides a ``benchmarking technique'' for estimating this parameter, which also does not provide any provable guarantee on the quality of the estimation of the parameters. Finally, SybilLimit can provide guarantees on the number of Sybil identities introduced per attack edge as long as $g=o(\frac{n}{\log n})$. Notice that both SybilGuard and SybilLimit do not require any global knowledge of the social network they operate on, and can be implemented in a fully decentralized manner.

\subsection{SybilInfer}
	SybilInfer uses a probabilistic model defined over random walks (called traces) in order to infer the the extent to which a set of nodes $X$, which generated such traces, is honest. The basic assumption in SybilInfer is that each node has a global view and knowledge of the social network, the network is fast mixing, and the node that initiates SybilInfer is a honest node. Techically, SybilInfer tries to ultimately label the different nodes in the graph into honest or Sybil nodes. In SybilInfer, each node, in a network of $n$ nodes performs $s$ walks, hence the overall number of walks in the universal trace is $s\times n$. Each trace among these traces consists of the first node (the initiator of the random walk) and the last node in the random walk (i.e., tail). Unlike the uniform (over node degree) transition probability used ins SybilGuard and SybilLimit, SybilInfer defines the transition matrix uniform over nodes, thus penalizing nodes with higher degree. The ultimate goal of the operation of SybilInfer is compute the probability $P(X=\text{Honest}|T)$; this is, computing the probability that a set of nodes $X$ being honest given the traces $T$. This probability is computed using Bayes theorem.

	SybilInfer also uses non-trivial techniques for sampling the honest configuration that is used initially for determining the honesty of a set of nodes from their traces. This sampling is performed using the Metropolis-Hasting algorithm by initially considering a set $X_0$ and modifying the set once at a time by either removing or adding nodes to the set: at each time, and with probability $P_{\text{add}}$ a new node $x$ from $\bar{X_0}$ is added to $X_0$ to make $X'=X_0\cup x$ or a node in $X_0$ is removed with probability $p_{\text{remove}}$. The process is performed for $n\log n$ rounds in order to obtain a good sample independent of $X_0$.

\subsection{SumUp}
	Unlike SybilGuard and SybilLimit which are generic to the problem of node admission, and decentralized in the sense that they do not require a single node to carry global information about the social graph, and SybilInfer which is applied for inferring honesty of nodes, SumUp~\cite{tran2009sybil} tries to tackle the Sybil attack in the context of vote aggregation.  In this context, a node---called the vote collector---wishes to collect votes in a Sybil resistant manner from other nodes in the network. This is, among a given number of votes on an object, the vote collector wishes to increase the fraction of votes accepted for honest nodes, reduce the accepted votes casted by the attacker through his attack edge and identify attackers once they misbehave repeatedly. At the core of SumUp, a link capacity assignment mechanism is used for adaptively assigning capacities to links in the trust-possessing social graph and for restricting the amount of votes propagated to the vote collector from the voters side. The adaptive vote flow mechanism of SumUp uses two observations of the conventional online voting systems: a few users in the system vote on a single object and that---if such voting system is implemented on top of a social graph---the congestion is only at links close to the vote collector. Accordingly, SumUp suggests to distribute a number of tickets on the different links in the social graph based on their distance (according to some levels computed using the breadth-first search algorithm) from the vote collector.

	One obvious draw of the technique is its high computational requirements: the running time of a typical algorithm such as the Ford-fulkerson algorithm would require an order of the number of edges of operations for collecting the vote of a single voter. The authors further suggest a heuristic that uses only an order of the graph diameter number of steps, where each node greedily selects a node at the higher level through which it is connected using a non-zero capacity and propagate the vote until it reaches the vote collector. At any time step, given that the greedy step may not result in a non-zero capacity, each node is allowed to explore other nodes for paths at the same or lower level.

\subsection{GateKeeper}
	GateKeeper~\cite{gatekeeper-infocom11} borrows tools from both of SumUp and SybilLimit for efficient operation. In particular, it tries to improve the performance of SybilLimit by incorporating the ticket distribution component of SumUp. Unlike in SumUp where nodes are admitted through a non-zero path from the voter to the collector, as explained earlier, GateKeeper only considers the ``ticket distribution'' phase of SumUp where tickets are used for admitting nodes by an admission controller. Such tickets are propagated from the controller to all nodes in the same way as in SumUp. However, in order to limit the attacker's chances of receiving more tickets and reduce his overall advantage, a controller in GateKeeper randomly selects $m$ different random nodes; called ``vantage nodes'', where a suspect node is admitted if and only if it receives $f_{\text{admit}}m$ tickets (where $f_{\text{admit}}$ is the fraction of randomly selected vantage points; $0.2$ is used in GateKeeper) from different vantage points. Therefore, a node is admitted if it is admitted by such fraction of he vantage points. To combat double spending, GateKeeper suggests the use of cryptographic signature chains of the paths through which the tickets are spent (propagated to the controller).

\subsection{Other Social Network-based DHTs}
SPROUT~\cite{MartiGG04} is another DHT routing protocol that uses social links of trust-possessing social graphs to route information to users operating on top of the social networks. SPROUT in fact builds on top of Chord~\cite{StoicaMKKB01} and adds additional links (routing table entries) in Chord to other users in the social network of any given node that are online at any time. By doing so, SPROUT claims to improve reliability and distribution of load of Chord itself.

Whanau was originally presented in~\cite{lesniewski2008sybil}, where the work in~\cite{lassK10} included further analysis and proof of performance and security as well as implementation and demonstration of the end-to-end guarantees. .

In~\cite{Danezis05sybil}, the authors use bootstrap graphs---trees that characterize introduction relationships in the DHT, in order to defend against the Sybil attack. By modifying the operation of Chord, the DHT of interest, in a way that each node returns address of all nodes that it knows of (including introduction points), the authors devise several strategies used to reduce the impact of Sybil attack. Unlike the original Chord which uses the closeness over the Chord as a metric for routing (query), the solution considers several strategies for routing, including diversity, mixed, and zig-zag. The authors show experimentally that such strategies can be used to more efficiently perform Sybil-resistant DHT lookups with less number of queries than that required by the plain Chord design, when operating under a Sybil attack. The design of MobID~\cite{QuerciaH10} is a social-network based Sybil defense that claims to provide a robust defense for mobile environments while existing defenses have largely been designed for peer-to-peer networks and are based on the random walk theory. Furthermore, MobID uses the \emph{betweenness}, a graph-theoretic measure in the social graph, for determining the goodness of nodes in order to defend against the Sybil attacks. The work, however, does not seem to provide any provable guarantees. A comparison between the various schemes, and others in the literature are in Table~\ref{tab:bounds} and Table~\ref{tab:boundsx}.

\begin{table*}[t]
  \begin{center}
  \caption{A comparison between the major schemes that use social graphs in terms of upper bounds on performance.  In all of these bounds, $O(\log n)$ is the mixing time of the social graph. The hidden constant here is very small; typically, $rn=0.005 n$. Model tunable parameters}\label{tab:bounds}
    \begin{tabular}{rrrrrrrr}
    \addlinespace
    \toprule
    {\bf Scheme}  & {\bf Maximum $g$} & {\bf Accepts} & {\bf $w$} & {\bf$\#$ walks}\\
    \midrule
    SybilGuard    &  $O(\sqrt{n}/\log n)^1$     & $O(\sqrt{n}\log n) $ & $O(\sqrt{n}\log n)$ & 1\\
    SybilLimit    &   $O(n/\log n)$    & $O(\log n)$ & $O(\log n)$ & $O(\sqrt{m})$\\
    SybilInfer      &   ---    &  --- & $O(\log n)$ & $c^3$\\
    SumUp	 	&$O(n)${$^2$}& $O(1)$ & ---&--- \\
    Gatekeepr	 	&$O(n/\log n)$ & $O(\log k)$ & $O(\log n)$ &$c^3$ \\
    Whanau	 	&$O(n/\log n)$ & $O(\log n)$ & $O(\log n)$ &$O(\sqrt{cn}\log n)$ \\
    MobID	 	& --- & --- & --- & --- \\
    \bottomrule
    \end{tabular}
    \end{center}
\end{table*}

\begin{table*}[top]
  \begin{center}
  \caption{A comparison between the major schemes that use social graphs in terms of their model, assumptions, and application.}\label{tab:boundsx}
    \begin{tabular}{rrrrr}
    \addlinespace
    \toprule
    {\bf Scheme} & {\bf Model}  & {\bf Graph assumptions} &  {\bf Application} & \\
    \midrule
    SybilGuard & Decentralized   &  Fast mixing/trust possessing & Generic\\
    SybilLimit & Decentralized   &   Fast mixing trust possessing & Generic\\
    SybilInfer & Centralized     &  Fast mixing/trust possessing & Generic\\
    SumUp	& Centralized 	&Fast mixing/trust possessing & Voting\\
    Gatekeepr	& Decentralized 	&Balanced expander (random)& Admission\\
    Whanau	& Decentralized 	&Fast mixing trust possessing& DHT\\
    MobID	& Decentralized 	& High betweenness honest nodes& Mobile\\
    \bottomrule
    \end{tabular}
    \end{center}
\end{table*}


\subsection{Recent analyses and supplementary work}
Our work in~\cite{miximc} initiated the study of the mixing time as the basic assumption used in Sybil defenses and showed negative results on its quality in many social networks. Viswanath et al. conducted an experimental analysis of sybil defenses based on social networks in~\cite{bimal}. Their study aimed at comparing different defenses (namely, SybilGuard~\cite{YuKGF08}, SybilLimit~\cite{YuGKX08}, SybilInfer~\cite{DanezisM09}, and SumUp~\cite{tran2009sybil}) independent of the data sets being used, by decomposing these defenses to their cores. They show that the different Sybil defenses work by ranking different nodes based on how well-connected are these nodes to a trusted node (the verifier). Also, they show that the different Sybil defenses are sensitive to community structure in social networks and community detection algorithms can be used to replace the random walk based Sybil defenses. We brought a similar insight in~\cite{mohaisen2011understanding} by showing that the core structure of social graphs is related to the mixing time, and utilized those findings in improving the mixing time of poorly-mixing ones in~\cite{mohaisenimproving}. We studied how trust affects the performance of Sybil defenses in~\cite{mohaisen2011keep}.



\section{Related works}\label{sec:related}
Related to our work, Levine et al \cite{levine2006survey} proposed a broad survey on solutions for sybil attack in general settings including P2P overlays. Unlike our work, they emphasized on classifying the literature works broadly rather than defining merits and shortcomings of each class of works. Our survey, however, has greatly benefited form their classification though the set of schemes reviewed in our survey is greatly different. In particular, the main technical contents of our survey review works that are published after the publication of the survey in  \cite{levine2006survey}. Related to social network-based defenses, Yu has presented an intriguing tutorial and a survey in~\cite{yu2011sybil}.

\section{Conclusion}\label{sec:conclusion}
The sybil attack is very powerful when applied to P2P overlays and their countermeasures are harder than in other networking settings because of the P2P overlays nature: centralized authorities necessary for security enforcement are discouraged and sometimes absent from P2P overlays designs. In this article, we review the literature of different methods used to defend against the Sybil attack in P2P overlays. We show the different defenses' assumptions, features, and shortcomings and compare them to each other.


\begin{thebibliography}{10}

\bibitem{Danezis05sybil}
George Danezis, Chris Lesniewski-laas, M.~Frans Kaashoek, and Ross Anderson.
\newblock Sybil-resistant dht routing.
\newblock In {\em In ESORICS}, Lecture Notes in Computer Science, pages
  305--318, Berlin, Heidelberg, 2005. Springer.

\bibitem{MaymounkovM02}
Petar Maymounkov and David Mazi{\`e}res.
\newblock Kademlia: A peer-to-peer information system based on the xor metric.
\newblock In Peter Druschel, M.~Frans Kaashoek, and Antony I.~T. Rowstron,
  editors, {\em IPTPS}, Lecture Notes in Computer Science, pages 53--65,
  Berlin, Heidelberg, 2002. Springer.

\bibitem{StoicaMKKB01}
Ion Stoica, Robert Morris, David~R. Karger, M.~Frans Kaashoek, and Hari
  Balakrishnan.
\newblock Chord: A scalable peer-to-peer lookup service for internet
  applications.
\newblock In {\em SIGCOMM}, pages 149--160, New York, NY, USA, 2001. ACM.

\bibitem{WangYL07}
Yong Wang, Xiao chun Yun, and Yifei Li.
\newblock Analyzing the characteristics of gnutella overlays.
\newblock In {\em ITNG}, pages 1095--1100, Washington, DC, USA, 2007. IEEE
  Computer Society.

\bibitem{PathakI06}
Vivek Pathak and Liviu Iftode.
\newblock Byzantine fault tolerant public key authentication in peer-to-peer
  systems.
\newblock {\em Computer Networks}, 50(4):579--596, 2006.

\bibitem{Douceur02thesybil}
John Douceur and Judith~S. Donath.
\newblock The sybil attack.
\newblock In {\em IPDPS}, pages 251--260, Washington, DC, USA, 2002. IEEE.

\bibitem{YuSKGX09}
Haifeng Yu, Chenwei Shi, Michael Kaminsky, Phillip~B. Gibbons, and Feng Xiao.
\newblock Dsybil: Optimal sybil-resistance for recommendation systems.
\newblock In {\em IEEE Symposium on Security and Privacy}, pages 283--298,
  Washington, DC, USA, May 2009. IEEE Computer Society.

\bibitem{CastroDGRW02}
Miguel Castro, Peter Druschel, Ayalvadi~J. Ganesh, Antony I.~T. Rowstron, and
  Dan~S. Wallach.
\newblock Secure routing for structured peer-to-peer overlay networks.
\newblock In {\em OSDI}, Berkeley, CA, USA, 2002. USENIX Association.

\bibitem{AdyaBCCCDHLTW02}
Atul Adya, William~J. Bolosky, Miguel Castro, Gerald Cermak, Ronnie Chaiken,
  John~R. Douceur, Jon Howell, Jacob~R. Lorch, Marvin Theimer, and Roger
  Wattenhofer.
\newblock Farsite: Federated, available, and reliable storage for an
  incompletely trusted environment.
\newblock In {\em OSDI}, New York, NY, USA, 2002. USENIX Association.

\bibitem{LedlieS05}
Jonathan Ledlie and Margo~I. Seltzer.
\newblock Distributed, secure load balancing with skew, heterogeneity and
  churn.
\newblock In {\em INFOCOM}, pages 1419--1430, Washington, DC, USA, 2005. IEEE.

\bibitem{PKIP2P}
Fran\c{c}ois Lesueur, Ludovic M{\'e}, and Val{\'e}rie Viet~Triem Tong.
\newblock An efficient distributed pki for structured p2p networks.
\newblock In {\em Proceeding of P2P}, pages 1--10, Washington, DC, USA, 2009.
  IEEE Computer Society.

\bibitem{LesueurMT08a}
Fran\c{c}ois Lesueur, Ludovic M{\'e}, and Val{\'e}rie Viet~Triem Tong.
\newblock A distributed certification system for structured p2p networks.
\newblock In David Hausheer and J{\"u}rgen Sch{\"o}nw{\"a}lder, editors, {\em
  AIMS}, volume 5127 of {\em Lecture Notes in Computer Science}, pages 40--52,
  Berlin, Heidelberg, 2008. Springer.

\bibitem{LesueurMT08b}
Fran\c{c}ois Lesueur, Ludovic M{\'e}, and Val{\'e}rie Viet~Triem Tong.
\newblock A sybil-resistant admission control coupling sybilguard with
  distributed certification.
\newblock In {\em WETICE}, pages 105--110, Washington, DC, USA, 2008. IEEE
  Computer Society.

\bibitem{LesueurMT08c}
Fran\c{c}ois Lesueur, Ludovic M{\'e}, and Val{\'e}rie Viet~Triem Tong.
\newblock A sybilproof distributed identity management for p2p networks.
\newblock In {\em ISCC}, pages 246--253, Washington, DC, USA, 2008. IEEE.

\bibitem{AvramidisKD07}
Agapios Avramidis, Panayiotis Kotzanikolaou, and Christos Douligeris.
\newblock Chord-pki: Embedding a public key infrastructure into the chord
  overlay network.
\newblock In Javier Lopez, Pierangela Samarati, and Josep~L. Ferrer, editors,
  {\em EuroPKI}, volume 4582 of {\em Lecture Notes in Computer Science}, pages
  354--361, Berlin, Heidelberg, 2007. Springer.

\bibitem{borisovPuzzle}
Nikita Borisov.
\newblock Computational puzzles as sybil defenses.
\newblock In Alberto Montresor, Adam Wierzbicki, and Nahid Shahmehri, editors,
  {\em Peer-to-Peer Computing}, pages 171--176, Washington, DC, USA, 2006. IEEE
  Computer Society.

\bibitem{YuGK07}
Haifeng Yu, Phillip~B. Gibbons, and Michael Kaminsky.
\newblock Toward an optimal social network defense against sybil attacks.
\newblock In Indranil Gupta and Roger Wattenhofer, editors, {\em PODC}, pages
  376--377. ACM, 2007.

\bibitem{YuGKX08}
Haifeng Yu, Phillip~B. Gibbons, Michael Kaminsky, and Feng Xiao.
\newblock Sybillimit: A near-optimal social network defense against sybil
  attacks.
\newblock In {\em IEEE Symposium on Security and Privacy}, pages 3--17,
  Washington, DC, USA, 2008. IEEE Computer Society.

\bibitem{YuKGF06}
Haifeng Yu, Michael Kaminsky, Phillip~B. Gibbons, and Abraham Flaxman.
\newblock {SybilGuard:} defending against sybil attacks via social networks.
\newblock In Luigi Rizzo, Thomas~E. Anderson, and Nick McKeown, editors, {\em
  SIGCOMM}, pages 267--278, New York, NY, USA, 2006. ACM.

\bibitem{YuKGF08}
Haifeng Yu, Michael Kaminsky, Phillip~B. Gibbons, and Abraham~D. Flaxman.
\newblock Sybilguard: defending against sybil attacks via social networks.
\newblock {\em IEEE/ACM Trans. Netw.}, 16(3):576--589, 2008.

\bibitem{gatekeeper-infocom11}
Nguyen Tran, Jinyang Li, Lakshminarayanan Subramanian, and Sherman~S.M. Chow.
\newblock Optimal sybil-resilient node admission control.
\newblock In {\em The 30th IEEE International Conference on Computer
  Communications (INFOCOM 2011)}, Shanghai, P.R. China, 2011. IEEE.

\bibitem{lassK10}
Chris Lesniewski-Lass and M.~Frans Kaashoek.
\newblock Wh\={a}nau: A sybil-proof distributed hash table.
\newblock In {\em 7th USENIX Symposium on Network Design and Implementation},
  pages 3--17, Berkeley, CA, USA, 2010. USENIX Association.

\bibitem{lesniewski2008sybil}
C.~Lesniewski-Laas.
\newblock {A Sybil-proof one-hop DHT}.
\newblock In {\em Proceedings of the 1st workshop on Social network systems},
  pages 19--24, New York, NY, USA, 2008. ACM.

\bibitem{SyversonGR97}
Paul~F. Syverson, David~M. Goldschlag, and Michael~G. Reed.
\newblock Anonymous connections and onion routing.
\newblock In {\em IEEE Symposium on Security and Privacy}, pages 44--54,
  Washington, DC, USA, 1997. IEEE Computer Society.

\bibitem{WangKAD}
Peng Wang, James Tyra, Eric Chan-tin, Tyson Malchow, Denis~Foo Kune, and
  Yongdae Kim.
\newblock Attacking the kad network, 2009.

\bibitem{levine2006survey}
B.N. Levine, C.~Shields, and N.B. Margolin.
\newblock {A survey of solutions to the sybil attack}.
\newblock Technical report, University of Massachusetts Amherst, Amherst, MA,
  2006.

\bibitem{RodrigruesLS02}
Rodrigo Rodrigues, Barbara Liskov, and Liuba Shrira.
\newblock The design of a robust peer-to-peer system.
\newblock In {\em 10th ACM SIGOPS European Workshop}, pages 1--10, New York,
  NY, USA, 2002. ACM.

\bibitem{NewsomeSSP04}
James Newsome, Elaine Shi, Dawn Song, and Adrian Perrig.
\newblock The sybil attack in sensor networks: analysis \& defenses.
\newblock In {\em IPSN '04: Proceedings of the 3rd international symposium on
  Information processing in sensor networks}, pages 259--268, New York, NY,
  USA, 2004. ACM.

\bibitem{FreedmanM02}
Michael~J. Freedman and Robert Morris.
\newblock Tarzan: a peer-to-peer anonymizing network layer.
\newblock In Vijayalakshmi Atluri, editor, {\em ACM Conference on Computer and
  Communications Security}, pages 193--206, Washington, DC, USA, 2002. ACM.

\bibitem{CornelliDVPS02}
Fabrizio Cornelli, Ernesto Damiani, Sabrina De~Capitani di~Vimercati, Stefano
  Paraboschi, and Pierangela Samarati.
\newblock Choosing reputable servents in a p2p network.
\newblock In {\em WWW}, pages 376--386, New York, NY, USA, 2002. ACM.

\bibitem{KangCLTKNWSHDK09}
Brent~ByungHoon Kang, Eric Chan-Tin, Christopher~P. Lee, James Tyra, Hun~Jeong
  Kang, Chris Nunnery, Zachariah Wadler, Greg Sinclair, Nicholas Hopper, David
  Dagon, and Yongdae Kim.
\newblock Towards complete node enumeration in a peer-to-peer botnet.
\newblock In Wanqing Li, Willy Susilo, Udaya~Kiran Tupakula, Reihaneh
  Safavi-Naini, and Vijay Varadharajan, editors, {\em ASIACCS}, pages 23--34,
  New York, NY, USA, 2009. ACM.

\bibitem{li2012sybilcontrol}
Frank Li, Prateek Mittal, Matthew Caesar, and Nikita Borisov.
\newblock Sybilcontrol: practical sybil defense with computational puzzles.
\newblock In {\em Proceedings of the seventh ACM workshop on Scalable trusted
  computing}, pages 67--78. ACM, 2012.

\bibitem{AhnBHL03}
Luis von Ahn, Manuel Blum, Nicholas~J. Hopper, and John Langford.
\newblock Captcha: Using hard ai problems for security.
\newblock In Eli Biham, editor, {\em EUROCRYPT}, volume 2656 of {\em Lecture
  Notes in Computer Science}, pages 294--311, Berlin, Heidelberg, 2003.
  Springer.

\bibitem{YanA07}
Jeff Yan and Ahmad Salah~El Ahmad.
\newblock Breaking visual captchas with naive pattern recognition algorithms.
\newblock In {\em ACSAC}, pages 279--291, Washington, DC, USA, 2007. IEEE
  Computer Society.

\bibitem{tran2009sybil}
N.~Tran, B.~Min, J.~Li, and L.~Subramanian.
\newblock {Sybil-resilient online content voting}.
\newblock In {\em NSDI}, Berkeley, CA, USA, 2009. USENIX.

\bibitem{MartiGG04}
Sergio Marti, Prasanna Ganesan, and Hector Garcia-Molina.
\newblock Dht routing using social links.
\newblock In {\em IPTPS}, pages 100--111, Washington, DC, USA, 2004. IEEE.

\bibitem{QuerciaH10}
Daniele Quercia and Stephen Hailes.
\newblock Sybil attacks against mobile users: friends and foes to the rescue.
\newblock In {\em INFOCOM'10: Proceedings of the 29th conference on Information
  communications}, pages 336--340, Piscataway, NJ, USA, 2010. IEEE Press.

\bibitem{miximc}
Abedelaziz Mohaisen, Aaram Yun, and Yongdae Kim.
\newblock Measuring the mixing time of social graphs.
\newblock In {\em Proceedings of the 10th annual conference on Internet
  measurement}, IMC '10, pages 383--389, New York, NY, USA, 2010. ACM.

\bibitem{bimal}
Bimal Viswanath, Ansley Post, Krishna~P. Gummadi, and Alan Mislove.
\newblock An analysis of social network-based sybil defenses.
\newblock In {\em SIGCOMM}, New York, NY, USA, August 2010. ACM.

\bibitem{DanezisM09}
George Danezis and Prateek Mittal.
\newblock {SybilInfer}: Detecting sybil nodes using social networks.
\newblock In {\em The 16th Annual Network \& Distributed System Security
  Conference}, Lecture Notes in Computer Science, Berlin, Heidelberg, 2009.
  Springer-Verlag.

\bibitem{mohaisen2011understanding}
Abedelaziz Mohaisen, Huy Tran, Nicholas Hopper, and Yongdae Kim.
\newblock Understanding social networks properties for trustworthy computing.
\newblock In {\em ICDCS Workshops}, pages 154--159. IEEE, 2011.

\bibitem{mohaisenimproving}
Abedelaziz Mohaisen and Scott Hollenbeck.
\newblock Improving social network-based sybil defenses by augmenting social
  graphs.
\newblock In {\em WISA}, 2013.

\bibitem{mohaisen2011keep}
Abedelaziz Mohaisen, Nicholas Hopper, and Yongdae Kim.
\newblock Keep your friends close: Incorporating trust into social
  network-based sybil defenses.
\newblock In {\em INFOCOM}, pages 1943--1951. IEEE, 2011.

\bibitem{yu2011sybil}
Haifeng Yu.
\newblock Sybil defenses via social networks: a tutorial and survey.
\newblock {\em ACM SIGACT News}, 42(3):80--101, 2011.

\end{thebibliography}

\end{document}